\documentclass[aps,prl,superscriptaddress,amssymb,showpacs,twocolumn]{revtex4-1}
\usepackage[english]{babel}
\usepackage[utf8]{inputenc}
\usepackage[T1]{fontenc}
\usepackage{graphicx}
\usepackage{amsmath}
\usepackage{hyperref}
\usepackage{makeidx}
\usepackage{comment}
\usepackage{ulem}
\usepackage{braket}
\usepackage{color}

\begin{abstract} 
We study the statistics of the kinetic (or equivalently potential) energy for $N$ non-interacting fermions in a $1d$ harmonic trap of frequency $\omega$, at finite temperature $T$. Remarkably, we find an exact solution for the full distribution of the kinetic energy, {\it at any temperature $T$ and for any $N$}, using a non-trivial mapping to an integrable Calogero-Moser-Sutherland model.  
As a function of temperature $T$, and for large $N$, we identify: (i) a quantum regime, for $T \sim \hbar \omega$, where quantum fluctuations dominate and (ii) a thermal regime, for $T \sim N \hbar \omega$, governed by thermal fluctuations. We show how the mean, the variance as well as the large deviation function associated with the distribution of the kinetic energy cross over from the quantum to the thermal regime as temperature increases.      
\end{abstract}

\begin{document}

\title{Kinetic energy of a trapped Fermi gas at finite temperature}

\author{Jacek Grela} \email{jacek.grela@lptms.u-psud.fr} 
\affiliation{LPTMS, CNRS, Univ. Paris-Sud, Universit\'{e} Paris-Saclay, 91405 Orsay, France}
\affiliation{M. Smoluchowski Institute
of Physics and Mark Kac Complex Systems Research Centre, Jagiellonian University, PL--30--059 Cracow, Poland}
\author{Satya N. Majumdar}
\email{majumdar@lptms.u-psud.fr}
\affiliation{LPTMS, CNRS, Univ. Paris-Sud, Universit\'{e} Paris-Saclay, 91405 Orsay, France}
\author{Gr\'{e}gory Schehr}
\email{gregory.schehr@u-psud.fr} 
\affiliation{LPTMS, CNRS, Univ. Paris-Sud, Universit\'{e} Paris-Saclay, 91405 Orsay, France}

\maketitle


The abundance of fermionic systems in nature makes them a fundamental subject to study in various areas, from astrophysics through atomic and nuclear physics, all the way to quantum information theory. The number of fermions $N$ spans many orders of magnitude with values typically varying from $N\sim 10^4-10^7$ in cold atoms~\cite{BDZ2008:COLDGASREV}, through $N\sim 10^2$ in complex nuclei~\cite{RS1980:NUCLEARBOOK} to $N\sim 10^1$ \cite{VSBYSC2001:QCOMP1,BCMCG2013:QCOMP2} in qubits. The characteristic temperature $T$ of these systems can also be vastly different, with cold atoms operating at $T\sim 10^{-7} K$ \cite{DMJ1999:FERMIGAS}, while neutron stars reaching $T\sim 10^{12} K$ \cite{G2000:NEUTRONSTAR}. The interactions between fermions in these systems also vary quite widely and in some systems, like in cold-atoms experiments, the interaction can even be tuned to almost zero to make them effectively non-interacting \cite{BDZ2008:COLDGASREV,GPS2008,MMT2010:QGAS2}.

\begin{figure}
\includegraphics[scale=.45]{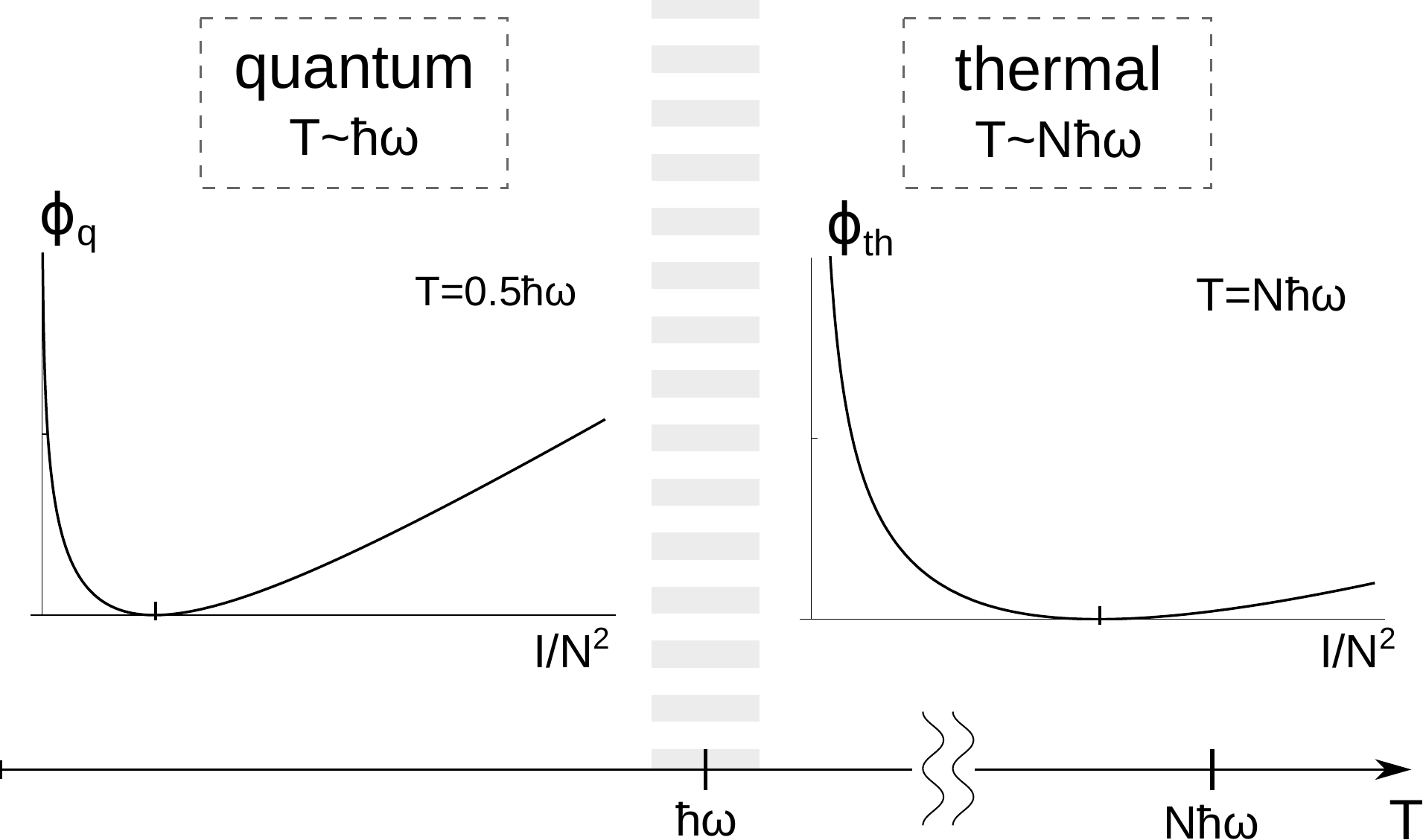}
\caption{Schematic sketch of the temperature axis with a clear separation between the quantum regime $T \sim \hbar \omega$ and the thermal regime $T \sim N \hbar \omega$, with the crossover regime shown as a shaded region. In each region, we also plot the rate functions $\phi_{\rm q}$ and $\phi_{\rm th}$ defined in Eqs. (\ref{large_dev_q}) and (\ref{def_rate_thermal}) respectively.
%
}
\label{fig1}
\end{figure}

However, even in this non-interacting limit, calculating the properties of various observables such as the fermion number fluctuations or the entanglement entropy of a subsystem are highly non-trivial, due to the strong repulsion (Pauli exclusion principle) between the fermions. This has been demonstrated in a number of recent articles \cite{SRLH2010,CMV2012,CC2004}. In addition, the presence of a confining trap (as in cold atom experiments) breaks the translational invariance and makes the problem even harder \cite{EIS2013:FERMIONS4,Vic2012,MMSV2014,CDM2015,DLMS2015:FERMIONS1,DLMS2015:EPL,DLMS2016:FERMIONS2}. While some results can be derived in the limit of large $N$ and 
at low or high temperatures, in general it is hard to find the full temperature and the $N$ dependence of these observables. In particular, it is important to study how the statistics of an observable cross over from the low $T$ limit (where quantum fluctuations are dominant) to the opposite high $T$ limit (where the system is governed mostly by thermal fluctuations). 
Hence it would be interesting  to find an experimentally accessible observable whose statistics can be computed analytically for all $N$ and at all temperatures~$T$. In this Letter we show that the statistics of the kinetic (or potential) energy of $N$ non-interacting fermions in a harmonic trap can be computed exactly for all $N$ and $T$. Our results demonstrate precisely how the quantum to thermal cross-over in the statistics of this observable
takes place as a function of temperature (see Fig. \ref{fig1}).

%
%
%

We consider the very simple system of $N$ spin-less non-interacting fermions in a $1d$ harmonic trap with
the Hamiltonian 
\begin{eqnarray}
\label{hamiltonian}
\hat{\mathcal{H}}_N = \sum_{i=1}^N \left[\frac{\hat{p}_i^2}{2m} +  \frac{1}{2} m \omega^2 \hat x_i^2 \right] \;.
\end{eqnarray}
With a trivial rescaling $\hat x_i \to \hat x_i/(m\omega)$, it is easy to see that $\hat x_i$ and $\hat p_i$ play a symmetric role in $\hat{\mathcal{H}}_N$. Consequently, the total kinetic energy $\hat E_K = \sum_{i=1}^N \hat p_i^2/(2m)$ and the total potential energy $\hat E_P = \sum_{i=1}^N  \hat x_i^2/(2m)$ (in the rescaled variables) are essentially the same observables. However, since they do not commute, they can not be measured simultaneously. For example, the potential energy $\hat E_P$ can be measured, in principle, by a quantum-gas microscope \cite{BGPFG2009:QGASMICROSCOPE,CNOGRBLZ2015:QGAS1} where the positions of individual particles are accessible experimentally. In contrast, the kinetic energy $\hat E_K$ should be accessible in the time-of-flight experiments \cite{MBKS1999:MOMENTAQGAS,P+2014:MOMENTAQGAS2} where the particles' momenta are measured typically. Hence $\hat E_P$, or equivalently $\hat E_K$, is a natural candidate for an experimentally accessible observable. In this Letter, we compute exactly, for any $N$, the quantum and thermal fluctuations of $\hat E_P$ (or equivalently that of $\hat E_K$).




This system of non-interacting fermions in a harmonic trap has been extensively studied in the recent past, both at $T=0$ \cite{EIS2013:FERMIONS4,MMSV2014} and finite $T>0$ \cite{DLMS2015:FERMIONS1,DLMS2016:FERMIONS2}. Several observables, including for instance the average density, the 
density-density correlation functions as well as the statistics of the rightmost fermion, have been studied in the large $N$ limit \cite{DLMS2015:FERMIONS1,DLMS2016:FERMIONS2}. At $T=0$ exactly, the many-body ground-state wave function is a Slater determinant of single particle harmonic oscillator wave functions and can be computed exactly \cite{EIS2013:FERMIONS4,MMSV2014}. The quantum probability distribution function (PDF) then reads 
\begin{align}
\label{mapping}
P_0({\bf x})=|\Psi_0({\bf x})|^2 = \frac{1}{{\cal Z}_0} \prod\limits_{i<j=1}^N |x_i - x_j|^2 e^{-\alpha^2 \sum\limits_{i=1}^N x_i^2},
\end{align}
where $\mathcal{Z}_0= \int |\Psi_0({\bf x})|^2 d{\bf x}$ with ${\bf x}=(x_1, \cdots, x_N)$ and the inverse length scale associated with the potential is
\begin{eqnarray}\label{def_alpha}
\alpha = \sqrt{m\omega/\hbar} \;.
\end{eqnarray}
At $T=0$, the fluctuations in the positions in Eq. (\ref{mapping}) are entirely due to quantum fluctuations. In addition, the $T=0$ quantum PDF in Eq. (\ref{mapping}) is identical to the joint distribution of the eigenvalues of a Gaussian Unitary matrix (GUE). At finite temperature $T\geq 0$, the corresponding joint PDF of the positions of the fermions reads
\begin{eqnarray}\label{PDF_finiteT}
P_T({\bf x}) = \frac{1}{{\cal Z}_T} \sum_{E} |\Psi_E({\bf x})|^2 e^{-\beta E},
\end{eqnarray} 
with $\mathcal{Z}_T = \sum_E \, e^{-\beta E}$, where $\beta = 1/T$ and $\Psi_E(\bf x)$ denotes a many-body eigenstate with energy $E$. The probability measure in Eq. (\ref{PDF_finiteT}) now encodes both quantum and thermal fluctuations. 
It is convenient to work in the position basis where the operator $\hat E_P$ is diagonal
\begin{equation}\label{EP_I}
E_P = \frac{1}{2}m\omega^2 I, \qquad {\rm where} \qquad  I = \sum_{i=1}^N x_i^2 \;.
\end{equation}
Henceforth, we use $I$ for convenience, instead of the potential energy $E_P$ (or the kinetic energy $E_K$). 

The PDF of $I$ is given by
\begin{equation}\label{PDF_I}
Q_T(I) = \left\langle \delta\left(I - \sum_{i=1}^N x_i^2 \right)\right\rangle_T  \;,
\end{equation}
where $\langle \cdots \rangle_T$ denotes an average with respect to the measure $P_T({\bf x})$ in Eq. (\ref{PDF_finiteT}). Its Laplace transform simply reads
\begin{eqnarray}\label{Q_T_laplace}
\tilde Q_T(p) = \left \langle e^{-p \sum_{i=1}^N x_i^2}\right \rangle_T \;.
\end{eqnarray}

\paragraph{$T=0$ case.} At $T=0$, the measure is given in Eq. (\ref{mapping}) where the normalization constant ${\cal Z}_0 \equiv \mathcal{Z}_0(\alpha^2)$ is only a function of $\alpha^2$. Moreover, this dependence can be simply obtained by rescaling $x_i \to \alpha x_i$ in Eq. (\ref{mapping}), which gives $\mathcal{Z}_0(\alpha^2) = \mathcal{Z}_0(1) \alpha^{-N^2}$. Consequently the Laplace transform $\tilde Q_0(p)$ in Eq. (\ref{Q_T_laplace}) with $T=0$ is immediately given by 
\begin{eqnarray}
\label{mhat0}
\tilde Q_0(p) = \frac{\mathcal{Z}_0(\alpha^2+p)}{\mathcal{Z}_0(\alpha^2)} = \left ( \frac{\alpha^2}{\alpha^2 + p} \right )^{N^2/2} \;.
\end{eqnarray}
Inverting this Laplace transform one obtains a gamma-distribution for $I = 2E_P/(m \omega^2)$
\begin{eqnarray}\label{Q0_exact}
Q_0(I) =  \frac{\alpha^{N^2}}{\Gamma(N^2/2)} e^{-\alpha^2I} I^{N^2/2-1} \;, \; I \geq 0 \;.
\end{eqnarray}
From Eq.~(\ref{Q0_exact}), the mean $\langle I\rangle_0$ and the variance ${\rm Var}(I)_0= \langle I^2\rangle_0 - \langle I\rangle_0^2$ are given by 
\begin{eqnarray}\label{mean_T0}
\langle I\rangle_0 = \frac{N^2}{2 \alpha^2} \;\;\; , \;\;\; {\rm Var}(I)_0 = \frac{N^2}{2\alpha^4}    \;.
\end{eqnarray} 
Note that in the large $N$ limit, $Q_0(I)$ in Eq. (\ref{Q0_exact}) can be expressed in the large deviation form 
\begin{eqnarray}\label{large_dev_T0}
Q_0(I) \sim e^{-N^2 \phi_{0}(I/N^2)} \;,
\end{eqnarray}
where $\phi_{0}(y) = \alpha^2 y - (1/2)(1+\ln\left(2\alpha^2\,y\right))$ is a convex rate function with a minimum at $y = \langle I\rangle_0/N^2 = 1/(2 \alpha^2)$.

\paragraph{The case $T \geq 0$ and finite $N$.} Unlike the $T=0$ case where the PDF $Q_0(I)$ is very simple to compute by exploiting the connection to GUE random matrices in Eq.~(\ref{mapping}), the corresponding PDF $Q_T(I)$ at finite $T$ is highly non trivial, as reflected in the complicated form of the quantum PDF $P_T({\bf x})$ in Eq. (\ref{PDF_finiteT}). Our main result in this Letter is to show that the Laplace transform $\tilde Q_T(p)$ in Eq. (\ref{Q_T_laplace}) can be obtained, for any finite $T$, as the ratio of two partition functions, albeit with two different temperatures
\begin{eqnarray}
\label{ratio}
\tilde{Q}_T(p) = \frac{\mathcal{Z}_{T'}}{\mathcal{Z}_T} \;,
\end{eqnarray}    
where ${\cal Z}_T = \sum_E e^{-\beta E}$ is the standard partition function at temperature $T = 1/\beta$ and ${\cal Z}_{T'} = \sum_E e^{-\beta' E}$ with $\beta' = 1/T'$. The effective temperature $T'$ is related to $T$ via a non trivial relation:
\begin{eqnarray}
\label{relation}
\cosh \hbar \omega \beta' = \cosh \hbar \omega \beta + \frac{p}{\alpha^2} \sinh \hbar \omega \beta \;.
\end{eqnarray} 
In the limit $T \to 0$, it is easy to see that one recovers the result in Eq. (\ref{mhat0}). 
%
This result in Eqs. (\ref{ratio}) and (\ref{relation}) is obtained by using a mapping of the harmonically confined {\it non-interacting} Fermi gas to an {\it interacting} Fermi
gas described by the Calogero-Moser-Sutherland (CMS) model~\cite{M1975:CMSSYSTEM,C1971:CMSSYSTEM,S1971:CMSSYSTEM} and using some combinatorial properties of the CMS model. Here, we briefly sketch the derivation of these formulae, relegating the details to the Supplementary Material (Supp. Mat.) \cite{GMS2017:SUPPLEMENTAL}. 

The mapping is achieved by transforming the Hamiltonian in Eq. (\ref{hamiltonian}) to a new CMS Hamiltonian. In the position basis, this transformation reads
\begin{eqnarray}
\tilde {\cal H}_N = - e^W({\cal H}_N - E_0)e^{-W},
\end{eqnarray}  
where $W= \frac{1}{4} m \omega^2 \sum_i x_i^2 - \sum_{j<k} \log |x_j - x_k|$ and $E_0 = \frac{\hbar \omega}{2} N^2$ is the ground-state energy of ${\cal H}_N$. The CMS Hamiltonian $\tilde {\cal H}_N$ can be written explicitly (see Supp. Mat. \cite{GMS2017:SUPPLEMENTAL}). The eigenfunctions of $\tilde {\cal H}_N$ are generalized Hermite polynomials labeled by Young tableaux whose properties have been studied recently~\cite{BF1997:CMSMODELPOLY,GP1999:CMSTEMP2,GGV2003:CMSTEMP1}. 
Using these properties, we can show (see Supp. Mat.~\cite{GMS2017:SUPPLEMENTAL}) that Eq. (\ref{ratio}) holds. Moreover, the partition function ${\cal Z}_T$ can be written explicitly as an $N$-fold product~\cite{GMS2017:SUPPLEMENTAL}    
\begin{eqnarray}
\label{z}
\mathcal{Z}_T = \prod_{i=1}^N \frac{e^{-\frac{\tau(i-1)}{2}}}{2\sinh \frac{\tau i}{2}} \;, \quad \tau = \hbar \omega \beta \;.
\end{eqnarray}
%
Below we first analyse the mean and the variance both for finite and large $N$.  

{\it Mean and variance}. By expanding $\tilde Q_T(p)$ in Eq. (\ref{Q_T_laplace}) around $p=0$, and using Eqs. (\ref{ratio}) and (\ref{relation})
one gets 
%
\begin{align}
\left < I \right >_T & = \frac{N(N-1)}{4\alpha^2} + \frac{1}{2\alpha^2} \sum_{k=1}^N k \coth \frac{\tau k}{2}, \label{mean}\\
\text{Var}(I)_T & = \frac{1}{2\alpha^4} \sum_{k=1}^N \frac{k^2}{\sinh^2 \frac{\tau k}{2} } + \frac{\coth \tau}{\alpha^2} \left < I \right >_T \;. \label{var}
\end{align}
While for the mean and the variance of $I$ we thus have explicit formulae for any $N$, its full distribution is hard to obtain explicitly for finite $N$ and $T>0$. However, in the large $N$ limit this is possible as will be shown below.   

We start by analysing the mean and the variance in Eqs. (\ref{mean}) and (\ref{var}) in the limit of large $N$. 
\begin{figure}
\includegraphics[scale=.5]{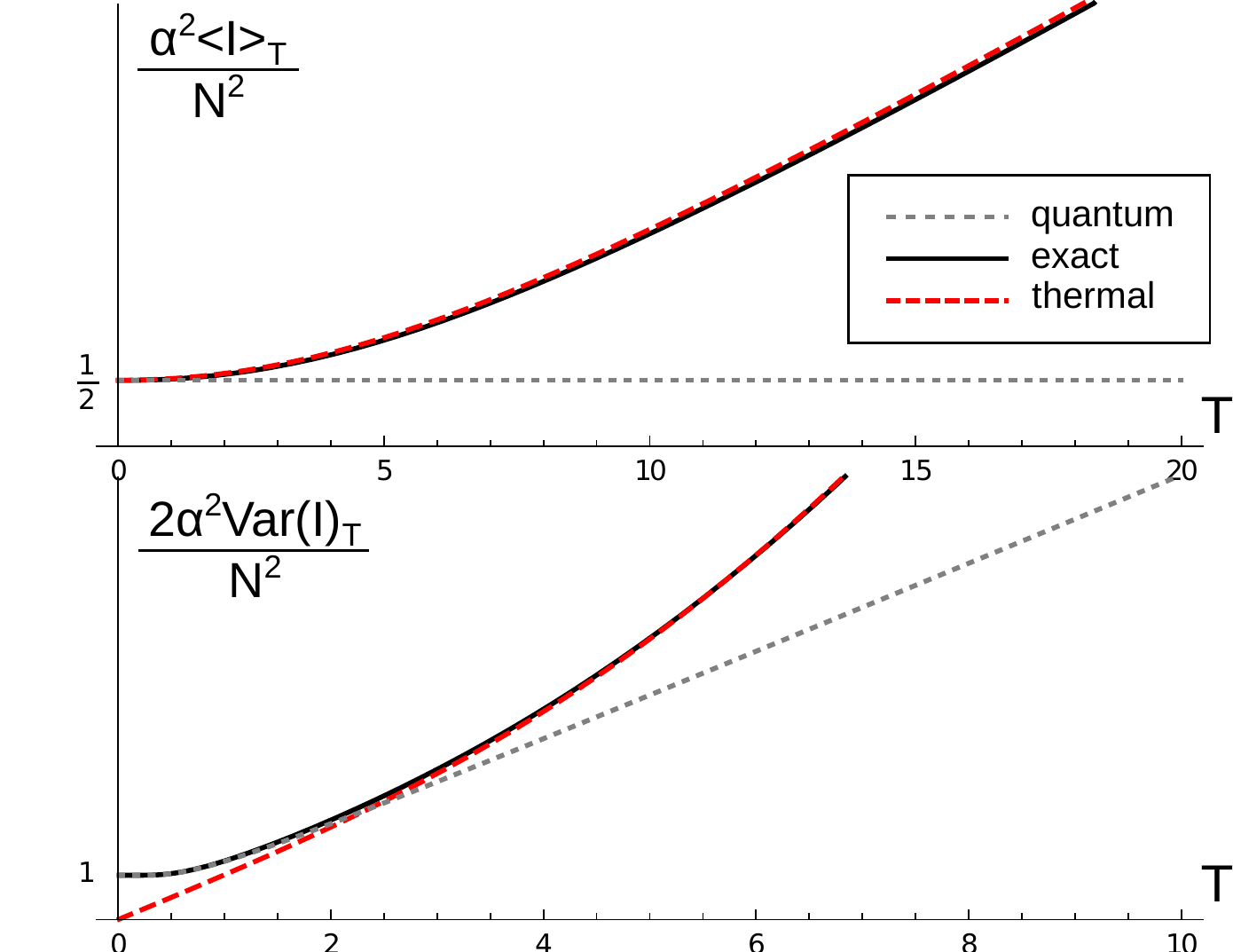}
\caption{(Color online) We plot the appropriately scaled mean (top panel) and the variance (bottom panel) of the potential energy, as a function of $T$ and for $N=30$. The solid black lines correspond to the exact results in Eqs. (\ref{mean}) and (\ref{var}), while the dotted (gray) and dashed (red) lines correspond to the large $N$ scaling results in the quantum and the thermal regimes respectively.
%
%
}
\label{fig2}
\end{figure}
To anticipate how temperature may affect the statistics of $I$, it is useful to go back to $T=0$, where the system is in the ground state. As discussed before, the Slater determinant characterizing the ground state wave function $\Psi_0(\vec x)$ consists of the lowest $N$ single particle wave functions, i.e., one fills up the first $N$ single particle energy levels up to the Fermi energy $E_F = (N-1/2)\hbar \omega$, with one fermion at each level. As one increases the temperature, one naturally encounters two temperature scales (see Fig. \ref{fig1} for a schematic representation). The first one corresponds to $T \sim \hbar \omega$, i.e., when the thermal energy is of order of the level spacing. In this case, only the fermions near the Fermi level get excited to higher levels. As $T$ increases further and becomes of the order of $T \sim E_F \approx N \hbar \omega$, all the fermions get affected and one essentially arrives at a fully thermal state. Therefore, when $T \sim \hbar \omega$ the system is sensitive to the discreteness of the spectrum and the quantum fluctuations are dominant. In contrast, for $T \sim E_F$, the thermal fluctuations take over the quantum fluctuations. We would expect that the large $N$ behavior of the mean and variance of $I$ will exhibit different behavior in these two temperature scales. Indeed our exact results in Eqs. (\ref{mean}) and (\ref{var}) demonstrate this explicitly.  

By analyzing Eq. (\ref{mean}) (see Supp. Mat. \cite{GMS2017:SUPPLEMENTAL}) we find for the mean of $I$
\begin{eqnarray}
\langle I \rangle_T
\sim 
\begin{cases}
& \frac{N^2}{\alpha^2} \, F_{\rm q}\left(\frac{T}{\hbar \omega}\right) \;,\;\;\;\;\, T\sim \hbar \omega, \\ 
 &\\
&\frac{N^2}{\alpha^2} \,F_{\rm th}\left(\frac{T}{N\hbar \omega}\right) \;,\; T\sim N \hbar \omega \;.
 \end{cases} 
 \label{meanapprox}
\end{eqnarray}
The scaling functions are given by
\begin{equation}
\label{meanapprox2}
F_{\text{q}}(u) = \frac{1}{2} \;, \quad F_{\text{th}}(z) =  -z^2\text{Li}_2\left (1-e^{1/z} \right) \;,
\end{equation}
where $\text{Li}_2(x) = \sum_{k=1}^\infty {x^k}/{k^2}$ is the dilogarithm function. While $F_{\rm q}(u)$ is trivially a constant function, $F_{\rm th}(z)$ is non trivial and has the asymptotic behaviors: $F_{\rm th}(z) \to F_{\rm th}(0)=1/2$ as $z \to 0$ and $F_{\rm th}(z) \sim z$ as $z \to \infty$. Thus, starting from the high temperature thermal scaling regime, if one takes $T \ll N \hbar \omega$, using $F_{\rm th}(0) =1/2$ one gets, ${\langle I \rangle_T} \sim N^2/(2\alpha^2)$. In contrast, starting from the low temperature quantum scaling regime, if one takes $T \gg \hbar \omega$, using $F_q(u) = 1/2$ (even as $u \to \infty$), we get ${\langle I\rangle_T} \sim N^2/(2 \alpha^2)$. This demonstrates an exact matching of the scaling behavior of the mean across the two scaling regimes. In Fig. \ref{fig2}, we plot $\alpha^2\langle I\rangle_T/N^2$ as a function of $T$, both for the exact result in Eq. (\ref{mean}) with $N = 30$ and the large $N$ scaling behavior corresponding to the quantum and the thermal regime.

Similarly, by analysing the variance of $I$ in Eq. (\ref{var}) in the large $N$ limit we get (see Supp. Mat. \cite{GMS2017:SUPPLEMENTAL})
\begin{eqnarray}
{\rm Var}(I)_T
\sim 
\begin{cases}
& \frac{N^2}{2\alpha^4} \, V_{\rm q}\left(\frac{T}{\hbar \omega}\right) \;,\;\;\;\;\, T\sim \hbar \omega, \\ 
 &\\
&\frac{N^3}{2\alpha^4} \,V_{\rm th}\left(\frac{T}{N\hbar \omega}\right) \;,\; T\sim N \hbar \omega \;.
 \end{cases} 
 \label{varapprox}
\end{eqnarray}
The two scaling functions are given by
\begin{eqnarray}
&&{V}_{\text{q}}(u) =\coth \frac{1}{u}, \label{varq} \\
&&{V}_{\text{th}}(z) = z\, \left ( 6 F_{\rm th}(z) - 1 - \coth\frac{1}{2z}  \right ) \;, \label{varth}
\end{eqnarray}
where $F_{\rm th}(z)$ is given in Eq. (\ref{meanapprox2}). The asymptotics of these scaling functions can be deduced easily. For example $V_{\rm q}(u) \sim 1 + 2\,e^{-1/(2u)}$ as $u \to 0$ and $V_{\rm q}(u) \sim u + 1/(3u)$ as $u \to \infty$. Similarly, $V_{\rm th}(z) \sim z$ as $z \to 0$ and $V_{\rm th}(z) \sim 4\,z^2+z/2$ as $z \to \infty$. Using these asymptotics, 
one can check, as in the case of the mean, that the variance in the quantum and in the thermal regime match smoothly as one increases the temperature. Indeed, for $T \gg \hbar \omega$ in the quantum scaling regime, one gets, using $V_{\rm q}(u) \sim u$ for large $u$, ${{\rm Var}(I)_T} \sim N^2 T/(2\alpha^4 \hbar \omega)$. Likewise, taking $T \ll N \hbar \omega$ in the thermal regime, and using $V_{\rm th}(z) \sim z$ for small $z$, we get the same result ${{\rm Var}(I)_T} \sim N^2 T/(2\alpha^4 \hbar \omega)$, thus ensuring a smooth matching of the variance. In Fig. \ref{fig2}, lower panel, we plot $2\alpha^4 {\rm Var}(I)_T/N^2$ as a function of $T$, both for the exact result in Eq. (\ref{var}) with $N=30$ and the large $N$ scaling predictions in Eq. (\ref{varapprox}).

We now turn to the full distribution $Q_T(I)$ [see Eq. (\ref{PDF_I})], whose Laplace transform is given in Eq. (\ref{ratio}). Inverting this Laplace transform, using Bromwich inversion formula, we obtain
\begin{eqnarray}\label{Brom_1}
Q_T(I) =  \frac{1}{2\pi i} \int_{\Gamma} dp \, e^{p\,I+ \ln \tilde Q_T(p)},
\end{eqnarray}
where $\tilde Q_T(p) = {\cal Z}_{T'}/{\cal Z}_T$ from Eq.~(\ref{ratio}) and the Bromwich contour $\Gamma$ is to the right of all singularities of the integrand. Using Eq.~(\ref{z}) one finds, $\ln \tilde Q_T(p) = - \frac{N(N-1)(\tau' - \tau)}{4} - \sum_{k=1}^N \log \frac{\sinh k \tau'/2}{\sinh k\tau /2}$, where
$\tau' = \hbar\omega \beta'$ and $\beta'$ is given in Eq. \eqref{relation}. We then analyse Eq.~(\ref{Brom_1}) in the large $N$ limit, using saddle point method. 

We start with the quantum regime where $T \sim \hbar \omega$. In this regime, $u = T/(\hbar \omega)$ is thus the natural scaling variable. For fixed $u$, Eq. (\ref{meanapprox}) shows that the mean $\langle I \rangle_T \sim N^2$, while Eq.~(\ref{varapprox}) predicts that the variance ${\rm Var}(I)_T \sim N^2$. For typical fluctuations around the mean on the scale of the standard deviation, one would expect from general central limit theorem a Gaussian form for $Q_T(I)$ with the above mean and variance. For larger atypical fluctuations, the Gaussian form no longer holds. Both the central Gaussian peak as well as the tails of $Q_T(I)$ are actually well described by a more general large deviation form 
\begin{eqnarray}\label{large_dev_q}
Q_T(I) \sim e^{-N^2 \phi_{\text{q}}\left (\frac{I}{N^2};u \right )} \;,
\end{eqnarray}
where $\phi_{\rm q}(y;u)$ is a rate function. For fixed $u$, as a function of $y$, $\phi_{\text{q}}\left (y;u \right)$ is expected to be a convex function, vanishing quadratically at the minimum at $y=y^* = \langle I \rangle_T/N^2$ [where $\langle I \rangle_T$ can be read off from the first line of Eq. (\ref{meanapprox})]. This quadratic form of the rate function near its minimum reproduces the Gaussian peak in $Q_T(I)$ around $I = \langle I \rangle_T$, with the correct $N$-dependent mean and the variance. By rescaling $p \to p/N$, and after a change of variable (see Supp. Mat. \cite{GMS2017:SUPPLEMENTAL}), one can reduce this integral in (\ref{Brom_1}) into a form which can be evaluated, for large $N$, by a saddle point method. Skipping details (see Supp. Mat. \cite{GMS2017:SUPPLEMENTAL}), we find that 
\begin{equation}
\label{phiq}
\phi_{\text{q}}(y;u) =  \frac{\text{sinh}^{-1} U-u^{-1}}{2} - \frac{\sqrt{1+U^2}-\sqrt{1+4 y^2 \alpha^4 U^2}}{2U},
\end{equation}
where $U = \frac{\sinh(1/u)}{2 y \alpha^2}$. We plot this function in the left panel of Fig. \ref{fig1}. Note that at $T=0$, i.e., $u=0$, the rate function $\phi_{\rm q}(y;0) = \phi_0(y)$ reduces to the zero temperature rate function given in Eq. (\ref{large_dev_T0}). 

We now switch to the thermal regime where $T \sim N \hbar \omega$. In this regime, $\langle I \rangle_T \sim N^2$ from Eq. (\ref{meanapprox}) and ${\rm Var}(I)_T \sim N^3$ from Eq. (\ref{varapprox}). For fixed $z = T/(N\hbar \omega)$, arguments similar to the quantum regime would suggest that $Q_T(I)$ has a large deviation form  
\begin{eqnarray}\label{def_rate_thermal}
Q_T(I) \sim e^{-N \phi_{\text{th}}\left (\frac{I}{N^2};z \right )} \;,
\end{eqnarray}
where $\phi_{\text{th}}\left (y;z \right)$ is the thermal rate function. Evaluating the integral over $p$ in Eq. (\ref{Brom_1}) by a saddle point method (similar to the quantum case), we can compute $\phi_{\rm th}(y;u)$. However, unlike in the quantum case, its expression is less explicit. We find~(see Supp. Mat. \cite{GMS2017:SUPPLEMENTAL})
\begin{eqnarray}
\phi_{\text{th}}(y;z) &=& \frac{1-v_* + y\alpha^2(v_*^2-1)}{2z} + \log \left ( \frac{\sinh\frac{1}{2z}}{\sinh\frac{v_*}{2z}} \right )+ \nonumber \\
&+&\frac{1}{z}\left(v_*F_{\rm th}\left(\frac{z}{v_*}\right)-F_{\rm th}(z)\right)
\label{phith}
\end{eqnarray}
where $F_{\rm th}(z)$ is given in Eq. (\ref{meanapprox2}) and $v_*$ is obtained implicitly, for given $y$ and $z$, by solving 
the equation $F_{\rm th}(z/v_*) = v_*\,y\alpha^2$. For a plot of this function $\phi_{\text{th}}(y;z)$, see the right panel of Fig.~\ref{fig1}. As in the case of the mean and the variance, one can check that the quantum $\phi_{\rm q}(y;u)$ as $u \to \infty$ matches with the thermal rate function $\phi_{\rm th}(y;z)$ as $z \to 0$.

In this paper, we have computed exactly the distribution of the kinetic/potential energy of $N$ non-interacting
fermions in a $1d$ harmonic trap, {\it at all temperatures and for any $N$}. Our results demonstrate explicitly how the
statistics of the kinetic energy cross over from the low temperature quantum regime ($T \sim \hbar \omega$) to the high
temperature thermal regime ($T \sim N \hbar \omega$). It would be interesting to investigate whether and how the present method (that relies on the mapping to an integrable Calogero-Moser-Sutherland model) can be extended to higher dimensions, $d>1$. In addition, our exact results rely crucially on the duality between the kinetic and the potential energy for the fermions in a {\it harmonic} trap. It will be challenging to investigate the effects of non-harmonic traps that break this duality, even though we would expect the existence of the two temperature scales even for non-harmonic traps.

JG acknowledges support from the National Science Centre, Poland under an agreement 2015/19/N/ST1/00878.




\newpage

\begin{widetext}

\begin{center}\Large{\textbf{Supplementary material}}\end{center}
\vspace*{1cm}
The Hamiltonian of a system of $N$ spinless fermions confined in a harmonic trap reads:
\begin{align}
\label{hamiltonian}
\mathcal{\hat H}_N = - \frac{\hbar^2}{2m} \sum_{i=1}^N \partial^2_{x_i} + \frac{1}{2} m \omega^2 \sum_{i=1}^N x_i^2 \;.
\end{align}
whose eigenvalue problem is given as
\begin{align}
\label{initeigen}
\mathcal{\hat H}_N \Psi_{E }(\textbf{x}) = E \,\Psi_{E} (\textbf{x})\;,
\end{align}
where $\Psi_{E} (\textbf{x})$ is the many-body eigenfunction with energy $E$. It can be constructed as a Slater determinant built from the single particle eigenfunctions $\phi_{n_i}(x_j)$
\begin{eqnarray} \label{end1}
\Psi_{E} (\textbf{x}) = \frac{1}{\sqrt{N!}}\det_{1\leq i,j \leq N} \phi_{n_i}(x_j)\;, \;\;\;\; {\rm with} \;\;\; \; n_1 > n_2 > \cdots > n_N \geq 0 \;,
\end{eqnarray}
where the corresponding energy, labelled by ${\bf n} = (n_1, \cdots, n_N)$, is given by
\begin{eqnarray}\label{enegy}
E = E_{\bf n} = \hbar \omega \sum_{i=1}^N ( n_i + 1/2) \;.
\end{eqnarray}
Note that $(n_1+1/2)\hbar \omega$ denotes the energy of the highest occupied single particle level. Similarly $(n_2+1/2)\hbar \omega$, $(n_3+1/2)\hbar \omega$, etc. denote the second, third, etc. highest occupied single particle levels. We can thus also label the many-body eigenfunctions by $\Psi_{\bf n} (\textbf{x})$. The normalized single particle eigenfunction corresponding to the level $n$ is given by  
\begin{align}
\phi_n(x) & = \sqrt{\frac{\alpha}{\sqrt{\pi}2^n n!}} e^{-\frac{\alpha^2 x^2}{2}} H_n(\alpha x)\;, \label{wavefd1}
\end{align}
with an inverse length-scale $\alpha = \sqrt{m\omega/\hbar}$. 

At zero temperature ($T=0$), the system is in its ground state. The ground state wave function $\Psi_0({\bf x})$ is obtained as the Slater determinant in Eq. (\ref{end1}) constructed from the first $N$ levels of the harmonic oscillator with ${\bf n} = (N-1,N-2,\cdots,0)$. In the ground state, the fermion positions fluctuate due to quantum fluctuations. Evaluating the Slater determinant explicitly, the quantum probability distribution function (PDF) reads
\begin{align}
\label{mapping_supp}
P_0({\bf x})=|\Psi_0({\bf x})|^2 = \frac{1}{{\cal Z}_0} \prod\limits_{i<j=1}^N |x_i - x_j|^2 e^{-\alpha^2 \sum\limits_{i=1}^N x_i^2}\;,
\end{align} 
where ${\cal Z}_0 = \int |\Psi_0({\bf x})|^2 d{\bf x}$ is the normalization constant. The ground state energy is given by the sum of the first $N$ single particle levels
\begin{eqnarray}\label{E0_supp}
E_0 = \hbar \omega \sum_{n=0}^{N-1} (n+1/2) = \frac{1}{2}{\hbar \omega N^2} \;.
\end{eqnarray}

At finite temperature, the system has quantum as well as thermal fluctuations which are encoded in the temperature dependent joint PDF 
\begin{eqnarray}\label{PDF_finiteT_supp}
P_T({\bf x}) = \frac{1}{{\cal Z}_T} \sum_{E} |\Psi_E({\bf x})|^2 e^{-\beta E} = \frac{1}{{\cal Z}_T} \sum_{n_1 > \cdots>n_N\geq 0} |\Psi_{\textbf{n}}(\textbf{x})|^2\,e^{-\beta E_{\textbf{n}}}  \;,
\end{eqnarray} 
with $\beta = 1/T$ and $\mathcal{Z}_T = \sum_E \, e^{-\beta E} = \sum_{n_1>\cdots>n_N\geq 0}e^{-\beta E_{\bf n}}$. The average value of any observable in the position basis $f({\bf x})$ is then given by
\begin{align}
\label{avtg02}
\left < f(\textbf{x}) \right >_T = \frac{1}{\mathcal{Z}_T} \sum_{n_1>\cdots>n_N\geq 0} e^{-\beta E_{\textbf{n}}} \int d\textbf{x} \,|\Psi_{\textbf{n}}(\textbf{x})|^2 f(\textbf{x}) \;.
\end{align}

\subsection{a) Derivation of Eqs. (12) and (13)}

We consider the potential energy in the position basis $E_P = (1/2)m\omega^2 I$ where $I=\sum_{i=1}^N x_i^2$. Its distribution is given by
\begin{align}
\label{defd=1}
Q_T(I) & = \left < \delta \left (I -  \sum_{i=1}^N x_i^2\right ) \right >_T \;,
\end{align}
where $\langle \cdots \rangle_T$ denotes an average with respect to the measure $P_T({\bf x})$ in Eq.~(\ref{PDF_finiteT_supp}). We denote its Laplace transform by 
\begin{eqnarray}\label{Q_T_laplace_sup}
\tilde Q_T(p) = \left \langle e^{-p \sum_{i=1}^N x_i^2}\right \rangle_T = \int_{0}^\infty  e^{-p\,I} \, Q_T(I) \, dI\;.
\end{eqnarray}

Our goal is to show that the Laplace transform $\tilde Q_T(p)$ in Eq. (\ref{Q_T_laplace_sup}) can be written as the ratio of two partition functions at two different temperature as in Eqs. (12) and (13) of the main text. From now on, we will work only in the position basis. To proceed, we follow Ref.~\cite{BF1997:CMSMODELPOLY_supp} and make the following transformation to the Hamiltonian ${\cal H}_N$ 
%
%
\begin{align}
\tilde{\mathcal{H}}_N = - e^W (\mathcal{H}_N - E_0) e^{-W}\;,
\end{align}
where $E_0$ is the ground state energy in Eq. (\ref{E0_supp}) and $W(\textbf{x}) = \frac{1}{4} m \omega^2 \sum_i x_i^2 - \sum_{j<k} \log |x_j - x_k|$. We read off from Ref.~\cite{BF1997:CMSMODELPOLY_supp} the transformed Hamiltonian:
\begin{align}\label{tilde_H_supp_1}
\tilde{\mathcal{H}}_N = \frac{\hbar^3 \omega}{4m} \sum_{i=1}^N \partial^2_{x_i} - \hbar \omega \sum_{i=1}^N x_i \partial_{x_i} + \frac{\hbar^3\omega}{2m} \sum_{k \neq j} \frac{1}{x_j - x_k} \partial_{x_j}\;,
\end{align}
which satisfies a modified eigenvalue equation
\begin{align}
\label{htilde}
\tilde{\mathcal{H}}_N \tilde{\Psi}_{\bm{\kappa}}(\textbf{x}) = \tilde{E}_{\bm{\kappa}} \tilde{\Psi}_{\bm{\kappa}} (\textbf{x})\;.
\end{align}
The eigenvalues and eigenvectors of \eqref{initeigen} and \eqref{htilde} are related by:
\begin{align}
\label{relation}
E_{\textbf{n}} & = E_0 - \tilde{E}_{\bm{\kappa}}\;, \qquad \Psi_{\textbf{n}}(\textbf{x}) = e^{-W(\textbf{x})} \tilde{\Psi}_\kappa(\textbf{x}) \;.
\end{align}
The transformed Hamiltonian in Eq. (\ref{tilde_H_supp_1}) corresponds to a Calogero-Moser-Sutherland type model. The eigenvalue equation in Eq. (\ref{htilde}) admits a family of solutions in terms of generalised Hermite polynomials $h_{\bm{\kappa}}$ that are labelled by ${\bm \kappa} = (\kappa_1, \cdots, \kappa_N)$ where $\kappa_i$'s are ordered $(\kappa_i \geq \kappa_{i+1})$ positive integers.   
The corresponding eigenfunction and eigenvalue read 
\begin{align}
\tilde{\Psi}_{\bm{\kappa}}(\textbf{x}) = h_{\bm{\kappa}}(\alpha \textbf{x})\;, \qquad \tilde{E}_{\bm{\kappa}} = - \hbar \omega |\bm{\kappa}| = - \hbar \omega \sum_{i=1}^N \kappa_i \;.
\end{align}
The sequence ${\bm{\kappa}}$ can be arranged in a Young tableau whose $i$-th row contains $k_i$ boxes (see Fig. \ref{fig1_supp}). Using \eqref{relation}, the eigenfunctions and eigenvalues of \eqref{initeigen} of the original Hamiltonian ${\cal H}_N$ can then be expressed as 
\begin{align}
\label{relation2}
\Psi_{\textbf{n}}(\textbf{x}) & = \frac{\alpha^{\frac{N^2}{2}}}{\sqrt{{\cal N}_{\bm{\kappa}}}} e^{-\frac{\alpha^2}{2} \sum\limits_{i=1}^N x_i^2} \prod_{j<k} |x_j - x_k| h_{\bm{\kappa}}(\alpha \textbf{x})\;, \qquad E_{\textbf{n}} = \frac{\hbar \omega}{2} N^2 + \hbar \omega |{\bm{\kappa}}| \;,
\end{align}
where $|{\bm{\kappa}}| = \sum_{i=1}^N \kappa_i$ and the constant
\begin{eqnarray}
{\cal N}_{\bm{\kappa}} = 2^{|{\bm{\kappa}}|} \Gamma(|{\bm{\kappa}}|+1) \, 2^{-N(N-1)/2}\, \pi^{N/2}\, \prod_{j=1}^{N} \Gamma(2+j) \label{N0_supp}
\end{eqnarray}
is chosen such that $\int d\textbf{x} |\Psi_{\textbf{n}}|^2 =1$.

We inspect the relation between the partitions $\bm{\kappa}$ and the excitations $\textbf{n}$. Recall that the ground state of the original Hamiltonian is labelled by ${\bf n}^{(0)}=(N-1,N-2,\cdots,0)$. For convenience, we denote $n_i^{(0)}=N-i$ for $i=1,2, \cdots, N$. Using this notation, we can then rewrite the energy eigenvalues of ${\cal H}_N$ in Eq. (\ref{enegy}) as 
\begin{align}\label{relation3}
E_{\textbf{n}} = \hbar \omega \sum_{i=1}^N (n_i + 1/2) = \frac{\hbar \omega}{2} N + \hbar \omega \sum_{i=1}^{N} n_i^{(0)} + \hbar \omega \sum_{i=1}^N (n_i - n_i^{(0)}) = \frac{\hbar \omega}{2} N^2 + \hbar \omega \sum_{i=1}^N (n_i - n_i^{(0)}) \;.
\end{align}
Comparing the energy eigenvalues in Eqs. (\ref{relation2}) and (\ref{relation3}) leads to the natural identification
\begin{align}
\label{ident}
\kappa_{i} = n_i - n_i^{0} = n_i +i -N\;, \qquad i=1,2, \cdots,N \;,
\end{align}
which then maps the excitations $\textbf{n}$ onto the Young tableau $\bm{\kappa}$. Since $n_i \geq N-i$ (recall that $n_1>n_2>\cdots>n_N\geq 0$), we have $\kappa_i \geq 0$ as well as $\kappa_{i}\geq \kappa_{i+1}$ for all $i=1, 2, \cdots, N$. We give an example in Fig. \ref{fig1_supp}.
\begin{center}
\begin{figure}[h]
\includegraphics[scale=.7]{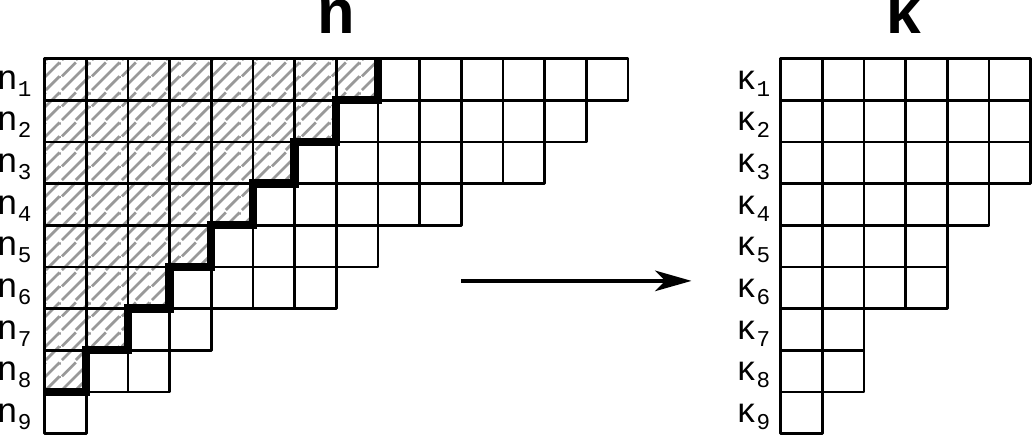}
\caption{An example of a mapping between the excitation vector $\textbf{n}$ and the Young tableau $\bm{\kappa}$ for $N=9$, $\textbf{n}=(14,13,12,10,8,7,4,3,1)$ and $\bm{\kappa} = (6,6,6,5,4,4,2,2,1)$. The shaded region is the ground state excitation $\textbf{n}^{(0)}=(8,7,\cdots,0)$.}
\label{fig1_supp}
\end{figure}
\end{center} 
We plug \eqref{relation2} into \eqref{avtg02} and reformulate the original problem of computing the averages as
\begin{align}
\label{fav}
\left < f(\textbf{x})\right >_T & = \frac{1}{\mathcal{Z}_T} t^{N^2/2} \sum_{\bm{\kappa}} t^{|\bm{\kappa}|}  \int d\textbf{x} \, |\Psi_{\textbf{n}}(\textbf{x})|^2 \, f(\textbf{x})\;,
\end{align}
where 
\begin{eqnarray}\label{def_t_supp}
t = e^{-\hbar \omega \beta}
\end{eqnarray}
and the partition function 
\begin{eqnarray}\label{def_Zt_supp}
\mathcal{Z}_T = t^{N^2/2} \sum_{\bm{\kappa}} t^{|\bm{\kappa}|} \;.
\end{eqnarray}
Introducing 
\begin{align}
\label{adef}
A(p) = (t\alpha^2)^{N^2/2} \int d\textbf{x} \, e^{- (p+\alpha^2) \sum\limits_{i=1}^N x_i^2} \, \prod_{j<k} |x_j - x_k|^2 \, \sum_{\bm{\kappa}} \frac{t^{|\bm{\kappa}|}}{\mathcal{N}_{\bm{\kappa}}} h_{\bm{\kappa}}(\alpha \textbf{x})^2\;, 
\end{align}
where ${\cal N}_{\bm{\kappa}}$ is given in Eq. (\ref{N0_supp}), it follows that $A(0) = \mathcal{Z}_T$ and in addition the Laplace transform in Eq. (\ref{Q_T_laplace_sup}) can be expressed as the ratio 
\begin{align}
\tilde{Q}_T(p) = \frac{A(p)}{A(0)} \;. \label{ratio_supp}
\end{align}
We next calculate $A(p)$ using an identity in Ref.~\cite{BF1997:CMSMODELPOLY_supp} 
that generalizes the Mehler kernel:
\begin{align}
\sum_{\bm{\kappa}} \frac{t^{|\bm{\kappa}|}}{\mathcal{N}_{\bm{\kappa}}} h_{\bm{\kappa}}(\alpha \textbf{x} )^2 = \frac{1}{\mathcal{N}_0} (1-t^2)^{-N^2/2} e^{-\frac{2t^2\alpha^2}{1-t^2} \sum\limits_{i=1}^N x_i^2} {}_0\mathcal{F}_0 \left ( \frac{2 \alpha t}{\sqrt{1-t^2}}\textbf{x}, \frac{\alpha}{\sqrt{1-t^2}} \textbf{x} \right ), \label{green}
\end{align}
where the generalized hypergeometric function ${}_0 \mathcal{F}_0 $ (with vector arguments) has the following expansion in terms of Jack polynomials $C_{\bm{\kappa}}({\bf x})$ (which are also labelled by Young tableaux):
\begin{eqnarray}
{}_0\mathcal{F}_0 \left ( \frac{2 \alpha  t}{\sqrt{1-t^2}}{\bf x}, \frac{\alpha}{\sqrt{1-t^2}}{\bf x} \right ) = \sum_{\bm\kappa} \frac{1}{|{\bm\kappa}|!} \frac{C_{\bm\kappa} \left (\frac{2 \alpha t}{\sqrt{1-t^2}}{\bf x} \right ) C_{\bm\kappa} \left (\frac{\alpha}{\sqrt{1-t^2}} {\bf x}\right )}{C_{\bm\kappa}({\bf 1})} &=& \sum_{\bm\kappa} \frac{1}{|{\bm\kappa}|!} \left ( \frac{2 t\alpha^2}{1-t^2} \right )^{|{\bm\kappa}|} \frac{C^2_{\bm\kappa} ({\bf x}) }{C_{\bm\kappa}({\bf 1})}  \nonumber \\
&=& \sum_{n=0}^\infty \frac{1}{n!} \left ( \frac{2t\alpha^2}{1-t^2} \right )^{n} \sum_{|{\bm\kappa}|=n} \frac{C^2_{\bm\kappa} ({\bf x}) }{C_{\bm\kappa}({\bf 1})} \;,\label{0f0}
\end{eqnarray}
where $|{\bm{\kappa}}| = \sum_{i=1}^N \kappa_i$ and ${\bf 1} = (1,1,\cdots,1)$. In these manipulations we use the fact that the Jack polynomials are homogeneous in their arguments. We plug formulae \eqref{green} and \eqref{0f0} to \eqref{adef} and find, after a rescaling of $x_i$'s:
\begin{align}
A(p) & = \frac{1}{\mathcal{N}_0} \left (\frac{t\alpha^2}{1-t^2} \right )^{N^2/2} \int d\textbf{x} \, e^{- \left (p+\alpha^2 \frac{1+t^2}{1-t^2} \right ) \, \sum\limits_{i=1}^N x_i^2} \prod_{j<k} |x_j - x_k|^2 \, {}_0\mathcal{F}_0 \left ( \frac{2 \alpha t}{\sqrt{1-t^2}}\textbf{x}, \frac{\alpha}{\sqrt{1-t^2}} \textbf{x} \right ) \nonumber \\
& = \sum_{n=0}^\infty \left ( \frac{t}{S_p(t)} \right )^{n+N^2/2} c_n\;, \label{Ap_supp_1}
\end{align}
where 
\begin{eqnarray}\label{Spt}
S_p(t) = \frac{p}{\alpha^2}(1-t^2) + (1 +t^2)
\end{eqnarray}
and the combinatorial coefficients 
\begin{eqnarray}\label{cn}
c_n = \frac{2^n}{\mathcal{N}_0 n!} \sum\limits_{|\bm{\kappa}|=n}\int d\textbf{y} \, \prod\limits_{j<k}|y_j - y_k|^2 \,e^{- \sum\limits_{i=1}^N y_i^2} \; \frac{C^2_{\bm{\kappa}} ( \textbf{y}) }{C_{\bm{\kappa}}(\textbf{1})}
\end{eqnarray}
are independent of both $t$ and $p$. Thus the partition function ${\cal Z}_T=A(0)$ is given by
\begin{eqnarray}\label{explicit_A0}
{\cal Z}_T \equiv {\cal Z}_T(t) = \sum_{n=0}^\infty \left(\frac{t}{1+t^2} \right)^{n+N^2/2}\,c_n\;, \;\quad \text{where} \;\quad t = e^{-\beta\hbar\omega} \;.
\end{eqnarray}
Furthermore, using Eq. (\ref{Ap_supp_1}) we can then express the Laplace transform in Eq. (\ref{ratio_supp}) as the ratio
\begin{align}
\tilde{Q}_T(p) = \frac{\sum\limits_{n=0}^\infty \left ( \frac{t}{S_p(t)} \right )^{n+N^2/2} c_n}{\sum\limits_{n=0}^\infty \left ( \frac{t}{1+t^2} \right )^{n+N^2/2} c_n}\;.
\end{align}
Since the denominator is just the partition function itself, it is natural to ask whether the numerator can be expressed as the same function ${\cal Z}_{T'}(t')$ but with an effective $t' = e^{-\beta'\hbar \omega}$ where $\beta' = 1/T'$. Indeed this can be done by expressing 
\begin{align}
\frac{t'}{1+t'^2} = \frac{t}{S_p(t)}\;,
\end{align}
which is solved by 
\begin{align}\label{tprime}
(t')_{\pm} = v \pm \sqrt{v^2-1}\;, \;\quad \text{where} \;\quad v = \frac{1}{2t}\left ( 1+ \frac{p}{\alpha^2} + t^2\left (1-\frac{p}{\alpha^2} \right ) \right ) \;.
\end{align}
Amongst the two roots, we choose $t'_-$ that ensures that as $p \to 0$, $t'\to t$, so that $\tilde Q_T(p \to 0) = 1$. This gives the Eq. (12) of the main text:
\begin{align}
\label{mhatfin}
 \tilde{Q}_T(p) = \frac{\mathcal{Z}_{T'}(t')}{\mathcal{Z}_{T}(t)}\;, \;\quad \text{where} \;\quad t' = v - \sqrt{v^2-1}\;.
 \end{align}
Finally, the new effective inverse temperature $\beta'$ can be related to $\beta$ by eliminating $v$ between $t' = v - \sqrt{v^2-1}$ and $v = \frac{1}{2t}\left ( 1+ \frac{p}{\alpha^2} + t^2\left (1-\frac{p}{\alpha^2} \right ) \right)$. Furthermore, using the identity, $\text{cosh}^{-1} v = -\log (v-\sqrt{v^2-1})$ valid for $v\ge 1$ gives our final result announced in Eq. (13) of the main text  
\begin{align}\label{betaprime_supp}
\cosh \hbar \omega \beta' = \cosh \hbar \omega \beta + \frac{p}{\alpha^2} \sinh \hbar \omega \beta \;.
\end{align}

\subsection{b) Derivation of Eq. (15)}

Here, we derive the expression for the partition function given in Eq. (15) of the main text. We start from the expression given in \eqref{def_Zt_supp} 
\begin{align}\label{generating_1}
\mathcal{Z}_T(t) = t^{N^2/2} \sum_{\bm{\kappa}} t^{|\bm{\kappa}|} \;,
\end{align}
where we recall that $|{\bm{\kappa}}| = \sum_{i=1}^N \kappa_i$ with $\kappa_{i} \geq \kappa_{i+1}$. Thus one can interpret $\mathcal{Z}_T(t)$ in Eq. (\ref{generating_1}) as the generating function of a counting problem:
\begin{align}\label{generating_2}
\mathcal{Z}_T(t) = t^{N^2/2} \sum_{M=0}^N p_M^{(N)} \, t^M \;,
\end{align}
where $p_M^{(N)}$ denotes the number of partitions of a positive integer $M \equiv |{\bm{\kappa}}|$ into at most $N$ parts. We find its generating function in Ref.~\cite{STA2011:ENUMCOMB_supp}
\begin{eqnarray}\label{generating_3}
\sum_{M=0}^N p_M^{(N)} \, t^M = \prod_{i=1}^N \frac{1}{1-t^i} \;.
\end{eqnarray}
Substituting this result in Eq. (\ref{generating_2}) gives the partition function
\begin{eqnarray}\label{generating_4}
\mathcal{Z}_T  = t^{N^2/2} \; \prod_{i=1}^N \frac{1}{1-t^i} \;.
\end{eqnarray}
Making further a change of variable $t = e^{-\tau}$ with $\tau = \beta \hbar \omega$, we get
\begin{align}\label{generating_5}
\mathcal{Z}_T = \prod_{i=1}^N \frac{e^{-\frac{\tau(i-1)}{2}}}{2\sinh \frac{\tau i}{2} }\;, \;\quad \text{where} \;\quad \tau = \beta \hbar \omega \;.
\end{align}
This gives the result in Eq. (15) of the main text.

\subsection{c) Derivation of Eqs. (16) and (17)}

We compute the first two moments of $I$ by expanding $\tilde{Q}_T(p)$ in Eq. (\ref{mhatfin}) up to $\mathcal{O}(p^2)$ around $p=0$. For this purpose, we substituted $v$ from Eq. (\ref{tprime}) in the expression of $t'$ given in Eq. (\ref{mhatfin}) and expanded $t'$ up to $\mathcal{O}(p^2)$. For the partition function $\mathcal{Z}_T (t)$, we use the expression in Eq.~\eqref{generating_4}. After some straightforward algebra, we get 
\begin{align*}
\tilde{Q}_T(p) & = 1 - \frac{{p}\, t }{\alpha^2}\frac{1}{\mathcal{Z}_T}\frac{\partial}{\partial t} \mathcal{Z}_T + \frac{p^2}{2\alpha^2}\frac{1}{\mathcal{Z}_T} \left ( t^2 \frac{\partial^2}{\partial t^2} \mathcal{Z}_T + \frac{2t}{1-t^2} \frac{\partial}{\partial t} \mathcal{Z}_T \right ) + \mathcal{O}(p^3) = \\
 & = 1 - p \left < I\right >_T + \frac{p^2}{2} \left < I^2 \right >_T + \mathcal{O}(p^3) \;.
\end{align*}
Comparing powers of $p$ on both sides, we get the mean and the variance as
\begin{align}
\left < I \right >_T & = - \frac{1}{\alpha^2 \hbar \omega} \partial_\beta \log \mathcal{Z}_T\;, \label{exact_mean_supp}\\
\text{Var}_T(I) & = \left < I^2 \right >_{\beta} - \left < I \right >_{\beta}^2 = \frac{1}{(\alpha^2 \hbar \omega)^2} \partial^2_{\beta} \log \mathcal{Z}_T - \frac{1}{\alpha^4 \hbar \omega}  \frac{1+t^2}{1-t^2} \partial_\beta \log \mathcal{Z}_T\;. \label{exact_var_supp}
\end{align}
Using the expression in Eq. (\ref{generating_5}), we get
\begin{align}
\label{freeZ}
\log \mathcal{Z}_T = -\hbar \omega \beta \frac{N(N-1)}{4} - \sum_{k=1}^N \log \left ( 2 \sinh \frac{\hbar \omega\beta}{2} k  \right ) \;.
\end{align}
Inserting this result in Eqs. (\ref{exact_mean_supp}) and (\ref{exact_var_supp}) we obtain
\begin{align}
\left < I \right >_T & = \frac{N(N-1)}{4\alpha^2} + \frac{1}{2\alpha^2} \sum_{k=1}^N k \coth \frac{\tau k}{2} \;, \label{Iav1}\\
\text{Var}_T(I) & = \frac{1}{2\alpha^4} \sum_{k=1}^N \frac{k^2}{\sinh^2 \frac{\tau k}{2} } + \frac{\coth \tau}{\alpha^2} \left < I \right >_T \;, \label{sigmaI1}
\end{align}
where $\tau = \beta \hbar \omega$. These expressions for the mean and the variance are exact for all $N$ and also for all temperature $T = 1/\beta$. This completes the derivation of Eqs. (16) and (17) of the main text.

\subsection{d) Derivation of Eqs. (18) and (19)}

We first analyse Eq. \eqref{Iav1} for the mean of $I$ in the large $N$ limit, both in the quantum ($T \sim \hbar \omega$) and in the thermal regime, $T \sim N \hbar \omega$.  

{\it Quantum regime.} We first set $u = T/(\hbar \omega) = 1/\tau$ and rewrite Eq. (\ref{Iav1}) as 
\begin{eqnarray}\label{meanI_1}
\left < I \right >_T &=& \frac{N(N-1)}{4\alpha^2} + \frac{1}{2\alpha^2} \sum_{k=1}^N k \left(\coth \frac{k}{2\,u} - 1 +1\right) \\
&=& \frac{N^2}{2\alpha^2} + \frac{1}{2\alpha^2} \sum_{k=1}^N k \left(\coth \frac{k}{2\,u} - 1\right) \;.
\end{eqnarray}
In the large $N$ limit, the discrete sum over $k$ converges to a constant (i.e., independent of $N$) and to leading order in the large $N$ we find
\begin{eqnarray}\label{meanI_2}
\left < I \right >_T = \frac{N^2}{2\alpha^2} + \mathcal{O}(1) \to \frac{N^2}{\alpha^2} F_{\rm q}\left(\frac{T}{\hbar \omega}\right)\;, \;\quad \text{where} \;\quad F_{\rm q}(u) = \frac{1}{2} \;.
\end{eqnarray}
This gives the result in the first line of Eq. (18) of the main text.

{\it Thermal regime.} In this regime, we set $z =T/(N \hbar \omega)$ in Eq. (\ref{Iav1}), or equivalently $\tau = 1/(Nz)$. In the large $N$ limit, the discrete sum can be replaced by a continuous integral using Euler-Maclaurin summation formula 
\begin{align}
\sum_{k=1}^N k \coth \left(\frac{k}{2Nz}\right) = N^2 \int_0^1 dx \, x \, \coth \frac{x}{2z} + \mathcal{O}(N)\;.
\end{align}
By performing this integral explicitly in terms of the dilogarithm function ${\rm Li}_2(x) = \sum_{k=1}^\infty x^k/k^2$, we obtain  
to leading order for large $N$
\begin{eqnarray}\label{mean_large_N_supp}
\left < I \right >_T = \frac{N^2}{\alpha^2} F_{\rm th}(z) + \mathcal{O}(N)\;, \;\quad \text{where} \;\quad F_{\rm th}(z) = -z^2 \text{Li}_2(1-e^{1/z}) \;.
\end{eqnarray}
Thus we get the second line of Eq. (18) of the main text.

\subsection{e) Derivation of Eqs. (20)-(22)}

We now analyse Eq. \eqref{sigmaI1} for the variance of $I$ in the large $N$ limit, respectively in the quantum ($T \sim \hbar \omega$) and in the thermal regime, $T \sim N \hbar \omega$.  

{\it Quantum regime.} We set $u = T/(\hbar \omega) = 1/\tau$ in Eq. (\ref{sigmaI1}) and take the large $N$ limit. In this case, the first term is of $O(1)$ since the sum $\sum_{k=1}^N k^2/\sinh^2(k/(2u))$ is convergent. Furthermore, in this regime $\langle I\rangle_T \sim N^2/(2\alpha^2)$. Hence to leading order for large $N$, we get  
\begin{align}\label{varq_supp}
\text{Var}(I)_{T} =  \frac{N^2}{2\alpha^4}V_{\rm q}\left(\frac{T}{\hbar \omega}\right) + \mathcal{O}(N)\;, \;\quad \text{where} \;\quad  V_q(u) = {\rm coth}(1/u) \;.
\end{align}
This gives the first line of Eqs. (20) and (21) of the main text. 

{\it Thermal regime.} In this regime, we set $z =T/(N \hbar \omega)$ in Eq. (\ref{sigmaI1}), or equivalently $\tau = 1/(Nz)$. In the large $N$ limit, the discrete sum can be replaced by a continuous integral using Euler-Maclaurin summation formula 
\begin{align}
\sum_{k=1}^N \frac{k^2}{\sinh^2{\left(\frac{k}{2Nz}\right)}} = N^3 \int_0^1 dx \, \frac{x^2}{\sinh^2{\left(\frac{x}{2Nz}\right)}}  + \mathcal{O}(N^2) \;.
\end{align}
In the second term of Eq.~(\ref{sigmaI1}), we have $\langle I \rangle_T \approx (N^2/\alpha^2)F_{\rm th}(z)$ from Eq.~(\ref{mean_large_N_supp}). Moreover, for large $N$, ${\rm coth}(1/(Nz)) \approx N\,z$ for $z \neq 0$ fixed. Thus both terms in Eq. (\ref{sigmaI1}) are of $\mathcal{O}(N^3)$ for large $N$. Collecting terms of $\mathcal{O}(N^3)$ together, we get 
\begin{eqnarray}
{\rm Var}(I)_T = \frac{N^3}{2 \alpha^4} V_{\rm th}\left(\frac{T}{N\hbar \omega}\right)\;, \qquad V_{\rm th}(z) = z \left(6 F_{\rm th}(z)-1-{\rm coth}\left(\frac{1}{z} \right) \right) \;, \label{varth_final_supp}
\end{eqnarray}
where $F_{\rm th}(z)$ is given in Eq. (\ref{mean_large_N_supp}). This completes the derivation of the second line of Eqs. (20) and (22) of the main text.

\subsection{f) Rate functions of Eqs. (25) and (27)}

We now inspect the full distribution function $Q_T(I)$ given by the Bromwich integral:
\begin{align}\label{start_QT_supp}
Q_T(I) = \frac{1}{2\pi i} \int_\Gamma dp \, e^{p\,I  + \ln \tilde{Q}_T(p)}\;,
\end{align}
with the Bromwich contour $\Gamma$ in the complex $p$-plane running to the right of all singularities of the integrand and
\begin{align}\label{lnQT_supp}
\ln \tilde Q_T(p) = -\frac{N(N-1)}{4} (\tau'-\tau) - \sum_{k=1}^N \ln \frac{\sinh(\frac{k\tau'}{2})}{\sinh(\frac{k\tau}{2})}\;, 
\end{align}
where $\tau=\hbar \omega \beta$, $\tau' = \hbar \omega \beta'$. In addition, $\tau'$ and $\tau$ are related by Eq. \eqref{betaprime_supp} that reads 
\begin{eqnarray}\label{tau_tauprime_supp}
\tau' = \text{cosh}^{-1} \left ( \cosh \tau + \frac{p}{\alpha^2} \sinh \tau \right) \;.
\end{eqnarray}

{\it Quantum regime.} We set $u = T/(\hbar \omega) = 1/\tau$ and take the large $N$ limit in Eq. (\ref{lnQT_supp}). We first rewrite
\begin{eqnarray}\label{identity_1}
\ln \frac{\sinh(\frac{k\tau'}{2})}{\sinh(\frac{k\tau}{2})} = \frac{k}{2}(\tau'-\tau) + \ln{\left(\frac{1-e^{-k\tau'}}{1-e^{-k\tau}} \right)} \;.
\end{eqnarray}
Then the sum in the second term is given by
\begin{eqnarray}\label{identity_2}
\sum_{k=1}^N \ln \frac{\sinh(\frac{k\tau'}{2})}{\sinh(\frac{k\tau}{2})} = \frac{N(N+1)}{4}(\tau'-\tau) + \sum_{k=1}^N \ln{\left(\frac{1-e^{-k\tau'}}{1-e^{-k\tau}} \right)} \;.
\end{eqnarray}
The second term in Eq. (\ref{identity_2}) is $O(1)$ since the sum is convergent. Using Eq. (\ref{identity_2}) in Eq. (\ref{lnQT_supp}) yields
\begin{eqnarray}\label{identity_3}
\ln \tilde Q_T(p) = - \frac{N^2}{2}(\tau'-\tau) + \mathcal{O}(1) \;.
\end{eqnarray} 
Using Eq. (\ref{tau_tauprime_supp}) for $\tau'$ we get
\begin{eqnarray}\label{identity_4}
\ln \tilde Q_T(p) = \frac{N^2}{2}\left[ \frac{1}{u} - \cosh^{-1}\left( \frac{p}{\alpha^2} \sinh{\frac{1}{u}}    + \cosh{\frac{1}{u}} \right) \right] + \mathcal{O}(1) \;.
\end{eqnarray}
Substituting this result (\ref{identity_4}) in Eq. (\ref{start_QT_supp}) and rescaling $I = y\,N^2$ we get
\begin{eqnarray}\label{identity_5}
Q_T(I) = \frac{1}{2\pi i} \int_\Gamma dp \, e^{N^2\, f(p)} \;,
\end{eqnarray}
where 
\begin{eqnarray}\label{identity6}
f(p) = y\,p+\frac{1}{2} \left[ \frac{1}{u} - \cosh^{-1}\left( \frac{p}{\alpha^2} \sinh{\frac{1}{u}}    + \cosh{\frac{1}{u}} \right) \right]\;,  \qquad {\rm where} \qquad y = \frac{I}{N^2} \;.
\end{eqnarray}
For large $N$, we evaluate the integral in Eq. (\ref{identity6}) by saddle point method. The saddle points are found from $f'(p)=0$, which has two solutions
\begin{align}
 p_\pm = \frac{\alpha^2}{\sinh 1/u} \left ( -\cosh 1/u \pm \sqrt{1+U^2} \right ) \;, \quad {\rm where} \quad   U = \frac{\sinh(1/u)}{2y\alpha^2} \;.
 \end{align} 
Since the Bromwich contour $\Gamma$ can be deformed to pass through the saddle $p_+$, we get to leading order for large $N$ 
\begin{align}\label{identity_7}
Q_T(I) \sim e^{-N^2\phi_{\text{q}}\left (\frac{I}{N^2};u \right )}\;,
\end{align} 
with the rate function $\phi_{\text{q}}(y;u) = - f(p_+)$ given explicitly by
\begin{align}
\phi_{\text{q}}\left (y;u \right ) & = \frac{\text{sinh}^{-1} U-u^{-1}}{2} - \frac{\sqrt{1+U^2}-\sqrt{1+4y^2\alpha^4 U^2}}{2U}\;, \quad {\rm where} \quad   U = \frac{\sinh(1/u)}{2y\alpha^2} \;. \label{identity_8}
\end{align}
This then provides the Eq. (25) of the main text.

{\it Thermal regime.} We set $z = T/(N \hbar \omega) = 1/(N \tau)$. We start from the Bromwich integral \eqref{start_QT_supp} and rescale the integration variable $p=q/N$ so that
\begin{align}
\label{identity_9}
Q_T(I) = \frac{1}{2\pi i N} \int_\Gamma dq \, e^{N \left ( q y +  \frac{1}{N} \ln \tilde{Q}_T(p=q/N) \right ) } \;, \qquad {\rm where} \;\quad y = \frac{I}{N^2} \;.
\end{align}
With the help of Eq. \eqref{identity_2} we rewrite the integrand:
\begin{align}
\label{identity_10}
\ln \tilde{Q}_T(p) = -\frac{N^2}{2} (\tau' - \tau) - \sum_{k=1}^N \ln \left ( \frac{1-e^{-k\tau'}}{1-e^{-k\tau}} \right )\;.
\end{align}
As a first step we expand the formula \eqref{tau_tauprime_supp} for $\tau'$ at $p=\frac{q}{N}$ and $\tau = \frac{1}{zN}$:
\begin{align}
\label{identity_11}
\tau' = \frac{v}{zN}  + \mathcal{O}(N^{-2})\;, \;\quad {\rm where} \;\quad v = \sqrt{1+ \frac{2qz}{\alpha^2}}\;,
\end{align}
and so the first term of Eq. \eqref{identity_10} reads
\begin{align}
\label{identity_12}
-\frac{N^2}{2} (\tau' - \tau) =  N\frac{1-v}{2z} + \mathcal{O}(1) \;.
\end{align}
The second term in Eq. (\ref{identity_10}) is computed using the Euler-Maclaurin summation formula
\begin{align}
\label{identity_13}
\sum_{k=1}^N \ln \left ( \frac{1-e^{-k\tau'}}{1-e^{-k\tau}} \right ) & = N \int_0^1 d\kappa \ln \left ( \frac{1 - e^{ - \frac{\kappa v}{z}}}{1-e^{-\frac{\kappa}{z}}}\right ) + \mathcal{O}(1) \nonumber \\
& = N \left [ \int_0^1 d\kappa \ln \left ( 1 - e^{ - \frac{\kappa v}{z}} \right ) - \int_0^1 d\kappa \ln \left ( 1-e^{-\frac{\kappa}{z}} \right ) \right ] + \mathcal{O}(1)\;,
\end{align}
where we have rescaled the integration variable by $k\to N \kappa$, plugged in Eq. \eqref{identity_11} and $\tau=\frac{1}{zN}$. Both integrals are of the type $\int_0^1 dx \ln \left ( 1-e^{-ax}\right ) = \frac{1}{a} \left ( \text{Li}_2(e^{-a})-\frac{\pi^2}{6} \right )$ with $a=v/z$ and $a=1/z$ respectively. Thus the second term of Eq. \eqref{identity_10} reads 
\begin{align}
\label{identity_14}
\sum_{k=1}^N \ln \left ( \frac{1-e^{-k\tau'}}{1-e^{-k\tau}} \right ) = N \left [ \frac{z}{v} \left ( \text{Li}_2 (e^{-v/z}) - \frac{\pi^2}{6} \right ) - z \left ( \text{Li}_2 (e^{-1/z}) - \frac{\pi^2}{6} \right ) \right ] + \mathcal{O}(1) \;.
\end{align}
Finally, we collect the results in Eqs. \eqref{identity_12} and \eqref{identity_14}, plug them into \eqref{identity_10} and compute the integrand of Eq.~\eqref{identity_9}:
\begin{align}
\label{identity_15}
qy +  \frac{1}{N} \ln \tilde{Q}_T(p=q/N) = qy + \frac{1-v}{2z} + \frac{z}{v} \left (\frac{\pi^2}{6} - \text{Li}_2 (e^{-v/z}) \right ) - z \left ( \frac{\pi^2}{6} - \text{Li}_2 (e^{-1/z}) \right ) + \mathcal{O}(1/N) \;,
\end{align}
where we recall that $v = \sqrt{1 + 2q z/\alpha^2}$. We now turn to calculating the integral \eqref{identity_9}. It is actually convenient to make a change of variable from $q$ to $v$ in the integration in Eq. (\ref{identity_9}). With this change of variable, this integral reads
\begin{align}
\label{identity_16}
 Q_T(I) = \frac{\alpha^2}{2\pi i N z} \int_{\Gamma'} dv \,v \,e^{ N f(v) }\;,
 \end{align} 
where $\Gamma'$ is the deformed contour of $\Gamma$ in the complex $v$-plane. The function $f(v)$ is given explicitly by
\begin{align}
\label{identity_165}
f(v) =  \frac{y\alpha^2}{2z}(v^2-1) +\frac{1-v}{2z} + \frac{z}{v} \left (\frac{\pi^2}{6} - \text{Li}_2 (e^{-v/z}) \right ) - z \left ( \frac{\pi^2}{6} - \text{Li}_2 (e^{-1/z}) \right ) \;,
\end{align}
where we have used $q = {\alpha^2}(v-1)/(2z)$. We recall that ${\rm Li}_2(x) = \sum_{k=1}^\infty x^k/k^2$ is the dilogarithm function. The integral in (\ref{identity_165}) can be evaluated, for large $N$, using saddle-point method. The saddle point equation $f'(v_*) = 0$ gives
\begin{align}
\label{identity_17}
-\frac{yv_*^3\alpha^2}{z^2} + \frac{v_*^2}{2z^2} + \frac{\pi^2}{6} + \frac{v_*}{z} \ln (1-e^{-v_*/z}) = \text{Li}_2(e^{-v_*/z})\;.
\end{align}
This formula simplifies considerably if we use the identity
\begin{align}
\label{identity_18}
\text{Li}_2(e^{-v/z}) =  \frac{v^2}{2z^2} + \frac{\pi^2}{6} + \frac{v}{z} \ln (1-e^{-v/z}) +\text{Li}_2(1-e^{v/z})\;,
\end{align}
which can be derived using Euler's reflection formula. The Eq. \eqref{identity_17} then simplifies to
\begin{align}
\label{identity_19}
 \text{Li}_2 \left (1-e^{v_*/z} \right ) = - \frac{yv_*^3\alpha^2}{z^2} \;.
\end{align}
Using $F_{\text{th}}(z) = -z^2\,\text{Li}_2(1-e^{1/z})$ from Eq. (\ref{mean_large_N_supp}), we can rewrite the saddle point equation as
\begin{align}
\label{identity_20}
F_{\text{th}}\left ( \frac{z}{v_*} \right ) = y v_* \alpha^2\;.
\end{align}
For $y>0$ and $z>0$, Eq. \eqref{identity_20} admits a real positive solution $v_* > 0$. Thus the saddle point solution finally reads
\begin{align}
\label{identity_21}
Q_T(I) \sim e^{-N \phi_{\text{th}}(y;z)}\;,
\end{align}
with $\phi_{\text{th}} (y;z) = -f(v_*)$:
\begin{align}
\label{identity_22}
\phi_{\text{th}}(y;z) = \frac{y\alpha^2}{2z}(v_*^2-1) +\frac{1-v_*}{2z} + \frac{z}{v_*} \left (\frac{\pi^2}{6} - \text{Li}_2 (e^{-v_*/z}) \right ) - z \left ( \frac{\pi^2}{6} - \text{Li}_2 (e^{-1/z}) \right )\;,
\end{align}
where $v_*$ is determined implicitly from Eq. \eqref{identity_20}, given $y$ and $z$. Using the identity in Eq.~\eqref{identity_18}, we can express the thermal rate function as
\begin{align}
\label{identity_23}
\phi_{\text{th}}(y;z) = \frac{1-v_* + y\alpha^2(v_*^2-1)}{2z} + \log \left ( \frac{\sinh\frac{1}{2z}}{\sinh\frac{v_*}{2z}} \right ) + \frac{v_*F_{\text{th}}(\frac{z}{v_*})-F_{\text{th}}(z)}{z}\;,
\end{align}
which results in Eq. (27) of the main text.

\end{widetext}


\begin{thebibliography}{10}

%
%

\bibitem{STA2011:ENUMCOMB_supp}
R.~P. Stanley,
\newblock {Enumerative Combinatorics}, volume~1 of {\it Cambridge Studies
  in Advanced Mathematics}.
\newblock Cambridge University Press, 2 edition, 2011.


\bibitem{BF1997:CMSMODELPOLY_supp}
T.~H. Baker and P.~J. Forrester,
\newblock {Comm. Math. Phys.}, {\bf 188}(1):175 (1997).

%
%
%
%
%
%
%
%
%
%
%
%
%
%
%
%
%
%
%
%
%
%
%
%
%
%
%
%
%
%
%
%
%
%
%
%
%
%



\end{thebibliography}
\end{document}